\documentclass{llncs}
\pdfoutput=1
\usepackage{makeidx}  % allows for indexgeneration
\usepackage[utf8]{inputenc}
\usepackage{amsmath,amsfonts}
\usepackage[lined,boxed,commentsnumbered]{algorithm2e}
\usepackage{graphicx}
\usepackage{subcaption}
\def\pik{\boldsymbol{\pi}^{(k)}}

\def\Sk{\boldsymbol{\Sigma}^{(k)}}

\def\Szt{\boldsymbol{\Sigma}^{(z_t)}}
\def\S0{\boldsymbol{\Sigma}_0}

\begin{document}

\title{Nonparametric Modeling of Dynamic Functional Connectivity in fMRI Data}
\titlerunning{Nonparametric Modeling Dynamic Functional Connectivity}  % abbreviated title (for running head)
\author{Søren F. V. Nielsen\inst{1} \and Kristoffer H. Madsen\inst{2} \and Rasmus Røge\inst{1} \and Mikkel N. Schmidt\inst{1} \and Morten Mørup\inst{1}}

\authorrunning{Nielsen et al.} % abbreviated author list (for running head)

%\institute{Anonymous Institute of Anonymity}
\institute{Section for Cognitive Systems, DTU Compute, Technical University of Denmark \\ 
Richard Petersens Plads, Building 324, DK-2800 Kgs. Lyngby \\
\email{sfvn@dtu.dk},\\
\and
Danish Research Centre for Magnetic Resonance, Section 714\\
Copenhagen University, Hospital Hvidovre, Kettegaard Allé 30, DK-2650 Hvidovre}

\maketitle              % typeset the title of the contribution

\begin{abstract}
 Dynamic functional connectivity (FC) has in recent years become a topic of interest in the neuroimaging community. Several models and methods exist for both functional magnetic resonance imaging (fMRI) and electroencephalography (EEG), and the results point towards the conclusion that FC exhibits dynamic changes. The existing approaches modeling dynamic connectivity have primarily been based on time-windowing the data and k-means clustering. We propose a non-parametric generative model for dynamic FC in fMRI that does not rely on specifying window lengths and number of dynamic states. Rooted in Bayesian statistical modeling we use the predictive likelihood to investigate if the model can discriminate between a motor task and rest both within and across subjects. We further investigate what drives dynamic states using the model on the entire data collated across subjects and task/rest. We find that the number of states extracted are driven by subject variability and preprocessing differences while the individual states are almost purely defined by either task or rest. This questions how we in general interpret dynamic FC and points to the need for more research on what drives dynamic FC.
\end{abstract}
\section{Introduction}
% Soft intro + Dynamic connectivty
The invention of functional magnetic resonance imaging (fMRI) paved the way for non-invasive studies of neuronal activity in the human brain. Especially the correlation of activity between brain regions, \emph{functional connectivity} (FC), has been of interest for over a decade. In recent years the term \emph{dynamic} functional connectivity has emerged in the field trying to explain temporal changes in the FC pattern \cite{hutchison2013dynamic,zalesky2014time,Calhoun2014-df}. Intuitively this makes sense since we expect that the interaction between segregated brain regions changes. These changes can be due to a number of factors such as the experimental design (task and resting state), non-experimental physical factors (fatigue, caffeine intake) and even neurological disorders \cite{calhoun2009functional}. 
An important aim of modeling FC as a dynamic concept is to run the models in a fully-unsupervised setting to achieve greater knowledge of how the brain functions and even extract biomarkers for diseases \cite{venkataraman2012whole}. One very prominent approach to analysing dynamic FC was presented in \cite{allen2012tracking}. Here windowed covariance matrices was extracted from a group-ICA representation of 405 healthy subjects resting state fMRI data. The upper triangular part of each extracted covariance matrix was then stacked into a vector yielding a vector space representation of the covariance structure for each window and subject. These vectors were then clustered using K-means clustering and the number of clusters was chosen using the Elbow-criterion. The results indicated that the resting state paradigm exhibits a temporally dynamical structure and that some of the FC patterns extracted diverge from the classical stationary results (such as the default mode network). In order to validate that the FC patterns (or states) extracted can be used to characterize resting state, we must test that the models find meaningful results in a setting where we have ground truth information available. \cite{Gonzalez-Castillo2015-ac} constructed an experiment where the participants were scanned while being asked to do different tasks in one continuous scan. They found that using windowed covariance matrices and k-means clustering (similar to \cite{allen2012tracking}) the cluster centroids could be used to explain the tasks carried out. But the number of dynamic states $k$ (fixed to the number of tasks) and window length had to be pre-specified whereas the model was not evaluated on independent test data.

In this paper we propose an extension of the infinite hidden Markov model (IHMM) \cite{beal2001infinite} tailored for the modeling of dynamic FC states in fMRI. We furthermore present a predictive likelihood framework for validating that the extracted structure can be used to characterize a motor task from rest. Using the IHMM circumvents having to specify the number of states in the model and window length. Instead the model can be viewed as an adaptive windowing method, where states can persist on different state specific time scales and the number of states learned as part of the inference. We will use this framework to investigate what drives dynamic FC.

\section{Methods}
\textbf{The IHMM-Wishart model:} A commonly used representation of FC is to represent the connectivity between areas of the brain by the covariance matrix. In a dynamic setting we model each state as having separate covariance matrices. We thus model fMRI in terms of a latent sequence of states, $z_t$ for $t=1..T$ that for each time point generates a brain image according to the state specific covariance matrix $\mathbf{x}_t\sim\mathcal{N}(\mathbf{0},\sigma_t^2\Szt)$, where $\sigma_t^2$ defines the magnitude of the state specific covariance structure $\Sk$ at time $t$ allowing states to be invariant to data magnitude but defined in terms of the connectivity profile. For the modeling of transitions between states we use the infinite hidden Markov model (IHMM), first proposed in \cite{beal2001infinite} and further developed in \cite{willsky2009nonparametric,van2012bayesian}. Completing the IHMM with a Gaussian distribution on the observed data and an inverse Wishart prior on the covariance structure, similar to the Wishart mixture modeling considered in \cite{hidot2010expectation,korzen2014quant} and the infinite Gaussian mixture model of \cite{Rasmussen1999-hd}, we arrive at the following generative model,
\begin{align}
\boldsymbol{\beta} &\sim \mathrm{GEM}(\gamma),  \label{eq:ihmm_gen_beta}\\
\pik|\boldsymbol{\beta} &\sim \mathrm{DP}(\alpha,\boldsymbol{\beta}), \label{eq:ihmm_gen_pi}\\
z_t | z_{t-1} &\sim \mathrm{Multinomial}(\boldsymbol{\pi}^{(z_{t-1})}), \label{eq:ihmm_gen_z}\\
\Sk &\sim \mathcal{W}^{-1}(\eta \S0, v_0), \label{eq:ihmm_wish_sk}\\
\mathbf{x}_t &\sim \mathcal{N}(\mathbf{0},\sigma_t^2\Szt)\label{eq:ihmm_wish_x}.
\end{align}
GEM is the stick-breaking construction (cf. \cite{sethuraman1994constructive,pitman2002poisson}), DP is the Dirichlet process, $\gamma$ and $\alpha$ are positive hyper-parameters controlling the state sequence, $\pik$ denotes the k'th row of the transition matrix $\boldsymbol{\pi}$, $\eta$ a is (positive) scale parameter, $\S0$ is a $p\times p$ matrix (covariance prior), and $\mathcal{W}^{-1}(\S0, v_0)$ denotes the inverse Wishart distribution, and $\sigma_t^2$ is the time specific covariance scaling. Viewing this in light of fMRI and FC (cf. also \cite{ryali2015variational,korzen2014quant}), each state can further be represented by a covariance matrix defining the FC where each state is invariant to magnitude of the FC due to the time-specific scaling parameter $\sigma_t^2$. We learn this parameter in the inference procedure (cf. next section) using a vague improper $1/\mathcal{X}$-prior. $\S0$ in the inverse Wishart prior plays the role of the default connectivity. In the experiments we estimate $\S0$ from a separate resting state fMRI scan. The parameter $\eta$ is the scaling or level of this default connectivity, which is learned during the inference. The degrees of freedom, $v_0$, set to the number of dimensions $p$ to make the inference well posed. 

\textbf{Inference:} For inference we use Markov chain Monte Carlo (MCMC). Due to conjugacy we can analytically integrate the state specific covariances $\Sk$ and infer the state sequence $\boldsymbol{z}$ using Gibbs sampling as described in \cite{van2012bayesian} and split-merge sampling \cite{jain2004split,hughes2012effective} which has been demonstrated to improved mixing and convergence properties for non-parametric Bayesian mixture models. We learn $\eta$ and $\sigma_t^2$ by Metropolis-Hastings with proposal distribution, $\eta^* = \exp(\ln \eta + z), z \sim
\mathcal{N}(0,\sigma = 0.1)$ (and similarly for $\sigma_t^2$) imposing an improper and uninformative $1/\mathcal{X}$-prior on both variables. The implementation of the model was done in MATLAB building on top of Juergen Van Gael's iHMM-toolbox \cite{iHMM-toolbox}. We used the code available online\footnote{\url{http://mloss.org/software/view/205/}} for sampling state-sequence relevant hyperparameters, where vague Gamma priors are placed on $\alpha$ and $\gamma$ and inferred via. an auxiliary variable Gibbs sampler \cite{teh2006hierarchical}. The predictive likelihood can be calculated using a modified Viterbi algorithm \cite{viterbi1967error}, and analytically integrating out $\sigma_t^2$ and using parameter posterior samples from the inference.

% \subsection{Predictive Likelihood}
% % Predictive Framework % What is the purpose?!
% In order to test if the model can capture specific task dynamics (and dynamics in general) we make use of the predictive likelihood utilizing posterior samples from the MCMC procedure. Thus by splitting our data into training and test we can evaluate how well a model trained on the training data $\mathcal{X}$ with parameters $\btheta$ predicts the test data $\mathcal{X}^*$  according to
% \begin{align}\label{eq:pred_like}
% p(\mathbf{X}^* | \mathbf{X}) = \int_{\btheta \in \mathcal{M}} p(\mathbf{X}^*| \btheta) p(\btheta| \mathbf{X}) d\btheta,
% \end{align}
% in which $\mathcal{M}$ is some model-defining space. This integral can be approximated by a modified version of the Viterbi algorithm \cite{viterbi1967error} using parameter posterior samples from $p(\btheta | \mathbf{X})$. This boils down to for each sample to sum over all time points' emission likelihood from all states weighted by their transition probability under the model, i.e. a posterior sample of the transition matrix $\boldsymbol{\pi}$. Note that we estimate a noise parameter for each time point in training, which we do not have available for the test data. Therefore we analytically integrate out $\sigma^2_t$ for the evaluation of the likelihood $p(\mathbf{X}^*|\btheta)$.

\section{Results}
We used fMRI data from a population of 29 subjects, that carried out a simple finger-tapping experiment (denoted motor) \cite{rasmussen2012model} as well as a resting state scan \cite{andersen2014non}. Preprocessing included normalization to MNI space and wavelet despiking \cite{Patel2014-ja}. We did a group ICA \cite{calhoun2001method} using the GIFT toolbox\footnote{\url{http://mialab.mrn.org/software/gift/}} into 20 components using the maximum likelihood ERBM algorithm \cite{li2010blind} with default settings. We discarded six components by visual inspection of the spatial maps representing common noise confounds. We split each subjects motor and resting-state data into a training (116 images) and a test set (120 images). Each training and test set was individually corrected for motion- and respiratory effects \cite{Friston1996-kk}, reference signal from cerebrospinal fluid (lateral ventricles) and white matter along with high-pass filtering (1/128 s cut-off) in a single regression step. The resting state scan was twice as long as the motor experiment so the second half of the resting state scan was removed and used for estimating the prior covariance $\S0$. For each of the 29 subjects training sets available (motor and rest) we ran ten IHMM-Wishart models and ten constant Wishart models, i.e. an IHMM-Wishart forced to be in one state. 

% Predictive Likelihood - Within subjects
First of all, we investigate if a model trained on the motor task (denoted a motor-model) predicts better on the motor-test set relative to a model trained on the resting state (denoted a rest-model). Second of all, we can test if predictive performance increases using a dynamic model, compared to a one-state constant model. For the motor data, we calculated the predictive likelihood for all ten runs, and made a paired t-test to evaluate if the predictive performance was better by the motor-model or the rest-model. The same was done for the resting state data. We saw on test data that within all 29 subjects, motor and rest was distinguishable with an IHMM-Wishart model. We furthermore tested if there was a significant difference between using a dynamic model vs. a static model. Except for one subject, this difference was not significant, which can be explained by the IHMM-Wishart finding almost exclusively one state on both motor and rest. 

% Group Analysis
The above analysis can be extended to group level, i.e. to investigate if a subjects task and resting state data can be distinguished by models trained on other subjects. We looked at 4 hypothesis tests, 1) that the motor- and rest-model predict equally well on motor data, 2)  that the rest- and motor-model predict equally well on resting state data, 3) that the motor-model is equal to using a one state constant model on motor data and 4) that the rest-model is equal to using a one state constant model on the resting state data. For motor  hypothesis 1 was rejected on average $26.5\pm 2$ out of $28$ times for a subject, and for resting-state hypothesis 2 was rejected on average $26.1 \pm 2$ out of $28$ times. Similarly as the within-subject analysis, very few dynamic models perform better on testing data than the one-state counterpart models. Hypothesis 3) and 4) could only be rejected $0.31 \pm 0.5$ and $0.24 \pm 0.6$ times respectively. 

\begin{figure}[t]
    \centering
    \begin{subfigure}[t]{\textwidth}
\centering
\includegraphics[width=\textwidth]{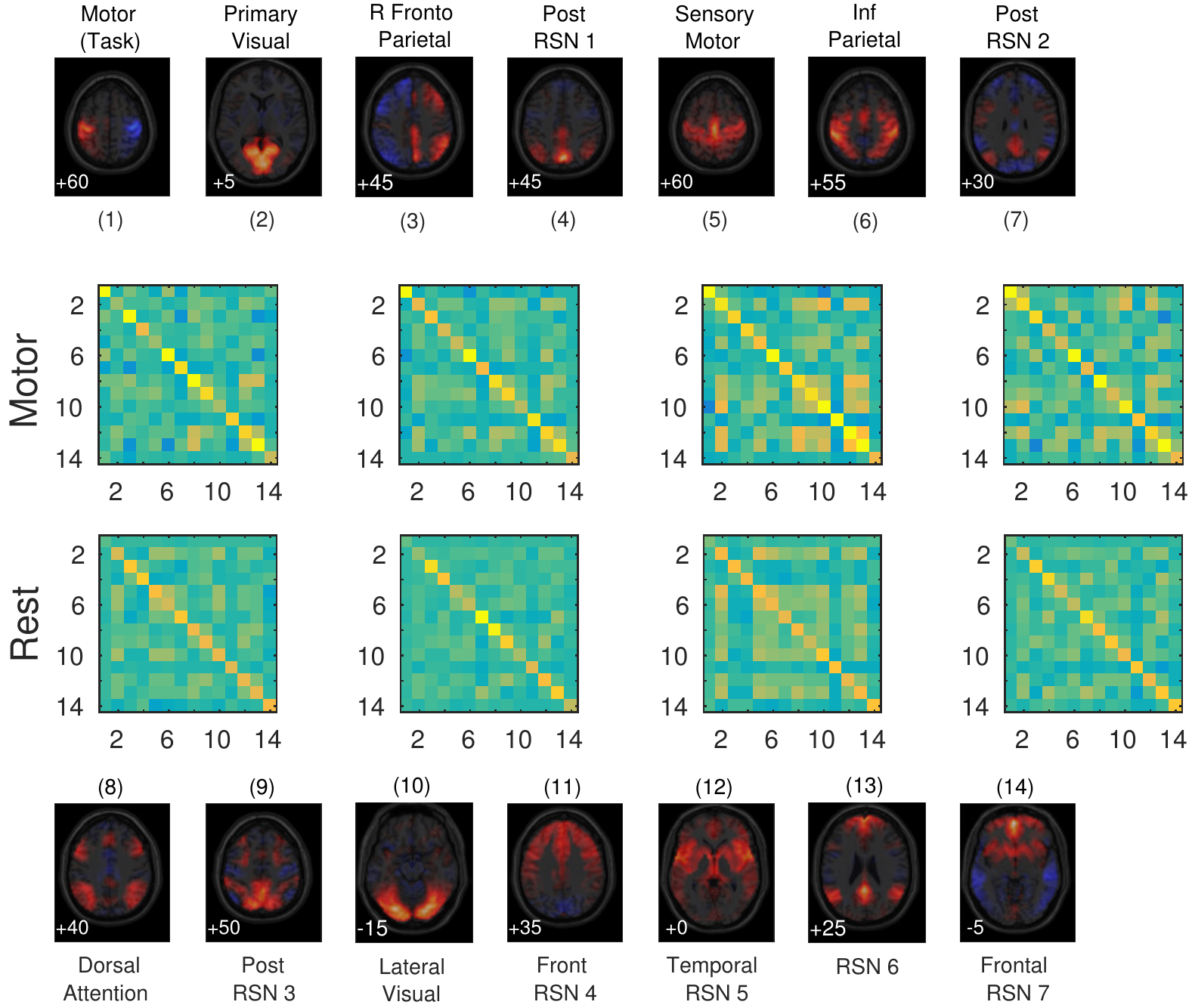}
\caption{Covariance estimate for each of the most populated states in the motor task (top row) and resting state (bottom row), along with the ICA component maps.}\label{fig:entirecol_pop}
\end{subfigure}

\begin{subfigure}[b]{0.49\textwidth}
	\includegraphics[height=4cm]{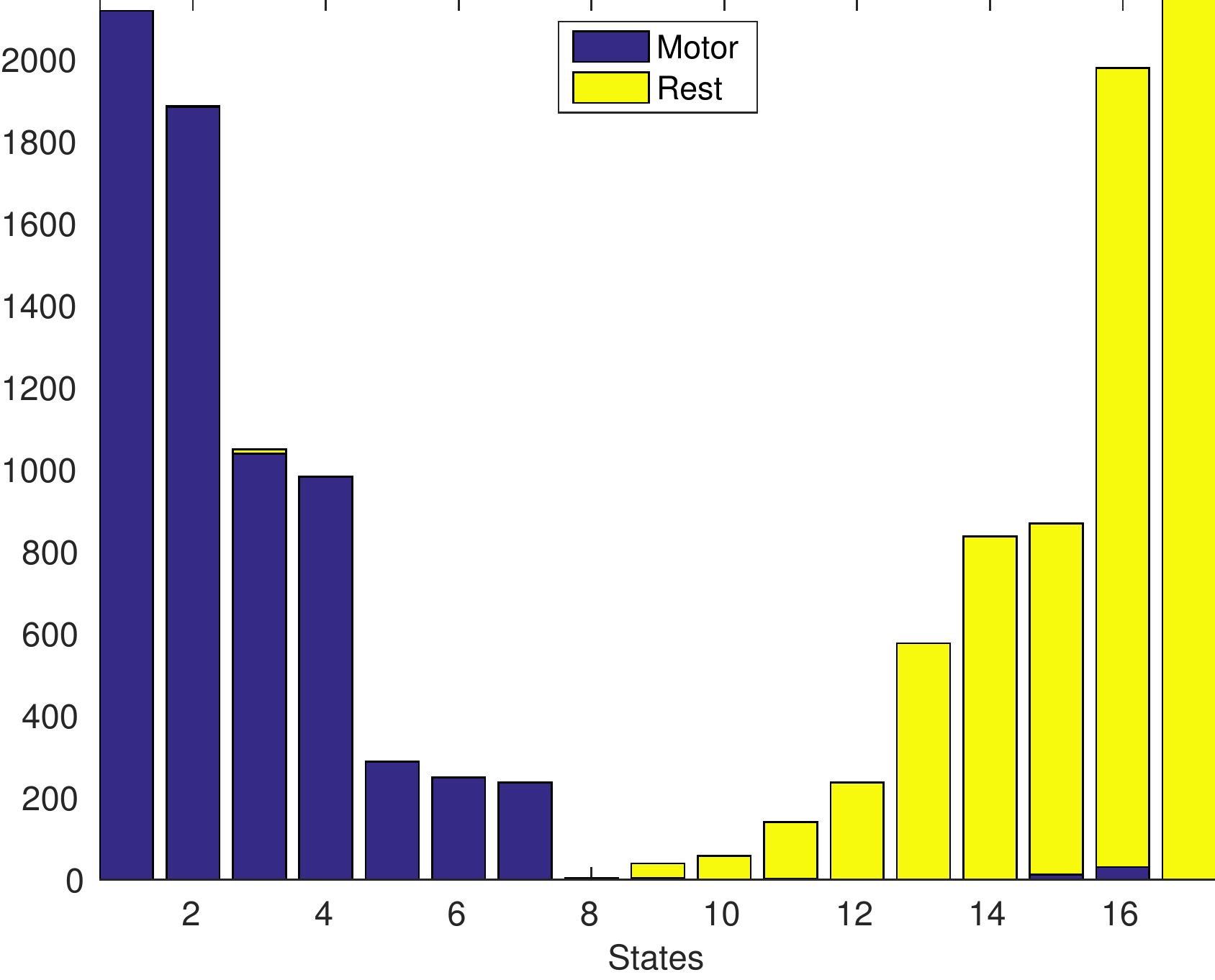}
	\caption{A bar plot displaying how frequently each state is populated. The states have been sorted by the difference in frequency in motor and rest-state.}\label{fig:entirecol_freq_z}
\end{subfigure}
\begin{subfigure}[b]{0.49\textwidth}
	\includegraphics[height=4.5cm]{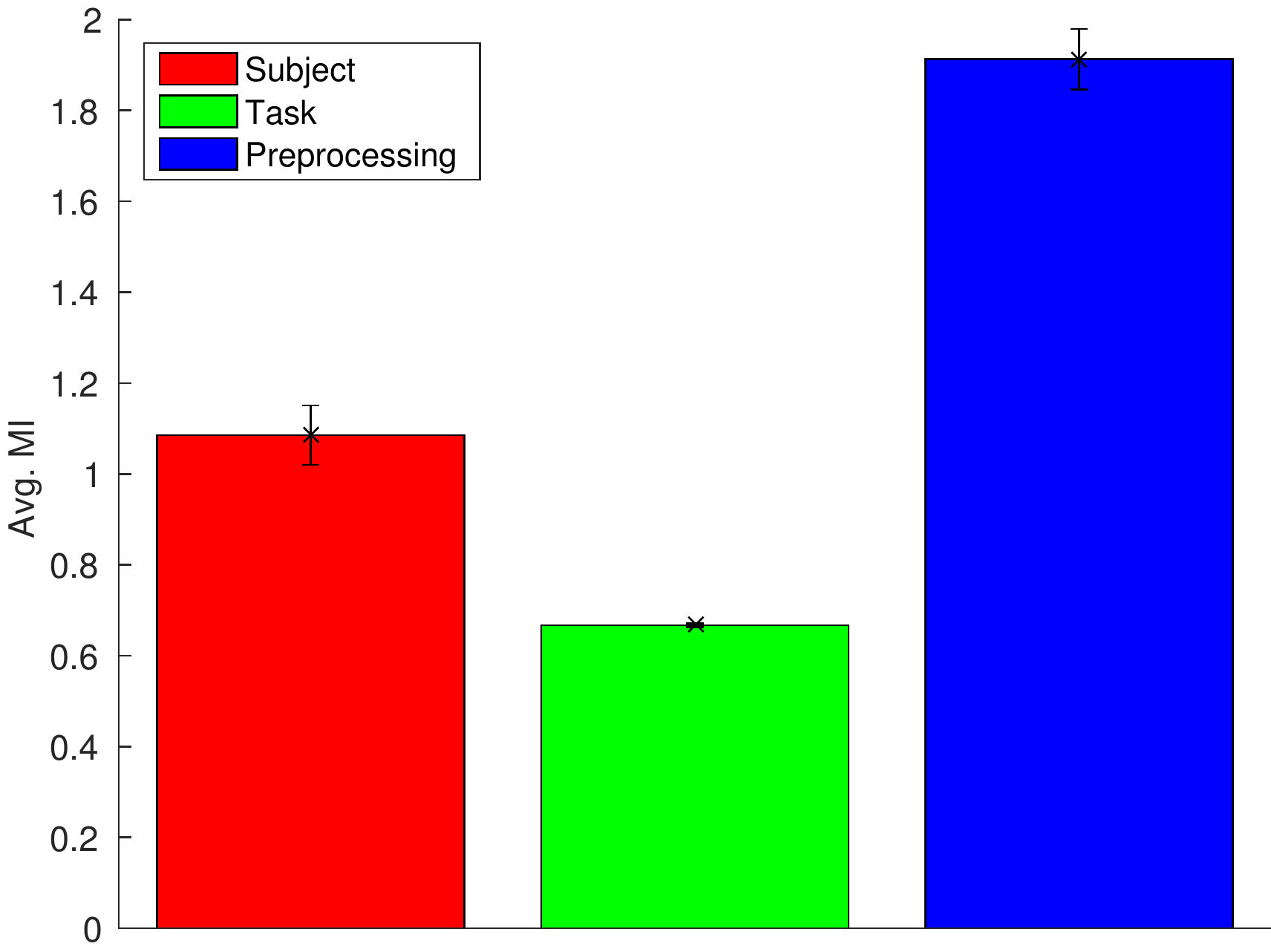}
	\caption{Average mutual information (MI) over samples with three state sequences, 1) subject (red), 2) task (green) and 3) preprocessing (blue)}\label{fig:entirecol_ami}
\end{subfigure}
 \caption{We collated all subjects motor and resting state experiments together and ran 5 chains of the IHMM-Wishart model. We show how the different states are populated, and investigate what drives the dynamics; subjects, tasks or preprocessing, for the sample with highest posterior probability.}\label{fig:entirecol}
\end{figure}

% Collated Analysis
To investigate dynamic FC at group level, we collated all subjects motor and resting state data together and ran the IHMM-Wishart fully unsupervised on the whole data set. In the modeling, states are shared across subjects and tasks but we handle discontinuities in the data by restarting the chain at each block that has been preprocessed individually (4 per subject).  We ran the IHMM-Wishart five times, and the sample with highest posterior probability was taken out for further analysis. The results can be seen in figure \ref{fig:entirecol}. In figure \ref{fig:entirecol_pop} we report the four most populated states' covariance matrices in both motor and resting-state along with the ICA components (IC). We note IC1 that well corresponds to the motor task is active in all the motor-states, and has been down-weighted in the resting-states. Inspecting the two tasks, the average number of states pr. subject in the motor-task was $3.28 \pm 0.21$ and for the resting state $3.60 \pm 0.27$. Furthermore, the states extracted are generally pertaining almost perfectly to either motor or rest (cf. figure \ref{fig:entirecol_freq_z}). However, from figure \ref{fig:entirecol_ami} we see that the average mutual information (MI) over posterior samples with state sequences describing the subjects, task and preprocessing shows that subjects and preprocessing drives the dynamics more than the tasks themselves.

\section{Discussion}
The IHMM-Wishart can distinguish between motor and rest on test data from the same subject. One confound is that the two experiments are recorded separately meaning that our ability to distinguish between motor and resting-state test data probably is caused by training and test set being recorded close in time. Since the group analysis shows that models trained on a subject can well distinguish between motor and rest of others, this confound is not present in this setting. This means that the IHMM-Wishart is able to well extract states that characterize motor from rest. In this analysis the model only found support for a single state in most of these single subject analyses carried out. The number of states found by the IHMM-Wishart is influenced though by the number of data points available and dimensionality, and this could influence the 'absence' of dynamics found due to the size of the training set (116 images). Collating data together from all subjects, we ran our model on the entire data set and saw each state were clearly pertaining to mainly either motor or rest. We see though that preprocessing has an impact on the states extracted - in an information theoretic perspective, the dynamics found are closer related to preprocessing than both to subjects and task (cf. figure \ref{fig:entirecol_ami}). 

% Final remarks
Our predictive framework demonstrates that the IHMM-Wishart can be used to characterize task vs. rest. This is a very controlled setting and should also be possible to achieve for simpler models than the one we are proposing here but evidences the utility of the IHMM-Wishart in characterizing fMRI data. Importantly, the IHMM-Wishart enables us to infer the number of dynamic states from data which for the collated data turned out to be \emph{more} than just two states, i.e. one for rest and one for motor. We here observed multiple task specific dynamic states driven by subject variability and preprocessing, which is aligned with the conclusion in \cite{korzen2014quant}. This tells us that care must be taken when interpreting dynamics at a group level as preprocessing even of blocks of data from the same subject evidences the presence of multiple dynamic functional states.

\clearpage

\section*{Acknowledgements}
This work was supported by the Lundbeck Foundation, grant no. R105-9813.

\bibliographystyle{unsrt}
\bibliography{dyna_brain.bib}

\end{document}